\documentclass[runningheads]{llncs}
\usepackage[T1]{fontenc}
\usepackage{graphicx}
\usepackage{tikz}
\usepackage{appendix}
\usepackage{amsfonts}
\usepackage{multirow}
\usetikzlibrary{fit,positioning,arrows.meta,backgrounds}
\usepackage{placeins}
\usepackage{xcolor}
\usepackage{soul}
\usepackage[normalem]{ulem}
\usepackage{hyperref}
\usepackage{enumitem}
\usepackage{amsmath}

\newcommand{\repeatthanks}{\textsuperscript{\thefootnote}}

\begin{document}
\title{A Width-Matched Comparison of Hybrid Quantum-Classical Self-Supervised Learning for Fingerprint Recognition}
\titlerunning{Hybrid SSL in Fingerprint Recognition}
\author{Maria S. Edwards\inst{1}\thanks{These authors contributed equally.}  \and
Kidwell Dlamini\inst{1}\repeatthanks \and
Pin-An Lin\inst{2}\thanks{These authors contributed equally.} \and
Wen-Hsien Hsu\inst{2}\repeatthanks \and
Wen-Chieh Fang\inst{1}\thanks{Corresponding author}
}
\authorrunning{Edwards et al.}
\institute{Department of Computer Science and Information Engineering, National Dong Hwa University, Hualien, Taiwan, \email{wcfang@gms.ndhu.edu.tw} \and
Department of Computer Science and Information Engineering, National Chiayi University, Chiayi City, Taiwan
}
\maketitle
\begin{abstract}
Fingerprint recognition is a widely deployed biometric, but supervised training requires large labeled enrollment sets. Self-supervised learning (SSL) removes this requirement, and hybrid quantum-classical models have been proposed to
enrich the learned representations. Prior quantum SSL studies consider a single contrastive objective, so it is unclear whether reported benefits depend on the objective or can be attributed to the quantum circuit. We insert the QuFeX
quantum feature-extraction module into three SSL frameworks, the contrastive SimCLR and MoCo v2 and the non-contrastive BYOL, and compare each hybrid with
its classical counterpart at matched representation width (8 features, equal to 8 qubits) on the SOCOFing fingerprint dataset, with a CIFAR-10 control, using
k-nearest-neighbor identification on encoder features. In single-run experiments the hybrid scores clearly higher for both contrastive objectives, whereas for BYOL a multi-seed analysis shows no reliable difference, suggesting
that any benefit depends on the SSL objective. A hardware-efficient circuit
(QNet) does not show the same gain. We examine whether the gains can be
attributed to the quantum circuit, considering circuit architecture, trainable
parameter count, nonlinearity, and the classical simulability of 8-qubit
circuits.
\end{abstract}

\section{Introduction}
Fingerprint recognition is among the most widely deployed biometric modalities,
used for device unlocking, access control, and identity verification, so high
recognition accuracy is important. Supervised approaches, however, require large
labeled enrollment sets that are costly and privacy-sensitive to collect.
Self-supervised learning (SSL) addresses this by learning representations from
unlabeled images: contrastive methods such as SimCLR \cite{simclr} and MoCo~v2
\cite{moco,mocov2}, and the non-contrastive BYOL \cite{byol}, have narrowed the gap with supervised
performance on natural images.
In parallel, hybrid quantum-classical neural networks have been proposed to enrich such representations\cite{hybrid,comparative,tunnelqnn}. Quantum Self-Supervised Learning (QSSL) inserts a parameterized quantum circuit into the encoder of a contrastive model, and prior
work reported that a quantum representation network can scores higher than a classical
one \cite{qssl}. Existing hybrid quantum-classical studies, however, are limited to a single contrastive objective \cite{qssl,qsea}, report a single or mixed accuracy
metric \cite{benckmark}, use single runs with large variance \cite{qrl}, and do not match the representation
width to the qubit count \cite{mnist}. Additionally, to our knowledge, quantum SSL has not been applied to fingerprint recognition, whose data have distinct characteristics. It therefore remains unclear whether any advantage is
(i) specific to the contrastive objective, (ii) present under settings where representation \emph{width} is matched, and (iii) attributable to the quantum circuit itself rather than to the additional classical parameters (encoding and residual connections) that accompany it, since matching width does not match module capacity.

This paper asks whether integrating a quantum feature-extraction module into self-supervised encoders yields a directionally consistent recognition benefit for fingerprints across different SSL objectives. We insert the QuFeX quantum module \cite{qufex} between the encoder and projection head of three SSL frameworks---SimCLR, MoCo~v2, and BYOL---and compare each hybrid model with its classical counterpart under a matched-width protocol (representation width $8$, equal to
$8$ qubits). 
We also test an alternative quantum circuit, QNet, on SimCLR as a secondary comparison; unlike QuFeX, QNet does not consistently scores higher the classical baseline at the single-layer depth used throughout our main experiments (Section~\ref{sec:results}, Appendix~\ref{sec:appendix}), which we use to motivate QuFeX as our primary hybrid module rather than to claim a general quantum advantage.
Recognition is evaluated with a $k$-nearest neighbor (KNN) monitor \cite{siamese,knn}, extracted from a publicly available implementation \footnote{https://github.com/zhirongw/lemniscate.pytorch}
on encoder features (Top-1 and Top-5) on the SOCOFing dataset \cite{sokoto}, with a CIFAR-10
control.

Our contributions are: 
\begin{enumerate}[label=(\arabic*)]
    \item a unified hybrid framework that inserts the same quantum module into contrastive and non-contrastive SSL encoders;
    \item a fair, width-matched classical-versus-hybrid comparison across three SSL objectives;
    \item the empirical finding that the hybrid QuFeX-based model scores higher than its classical counterpart on KNN recognition across all three objectives and on a CIFAR-10 control, though the margins are modest, obtained from single runs without variance estimates, and come at higher compute cost; a secondary QNet-based comparison does not show the same advantage at the only budget tested (15 epochs). (Section~\ref{sec:results}).
\end{enumerate}

This paper is organized into seven main sections. First, the introduction outlines the motivation, main objectives, and contributions of the paper. Second, the related works are presented and described. Third, the methodology describes the architecture and structure of the models utilized, as well as the evaluation methods used. Fourth, the experimental settings detail the setup, preprocessing steps, and integration. Fifth, the results obtained from the experiments are presented and explained. Sixth, the discussion aims to explain the reasoning behind the results when using different quantum architectures, the observations obtained after training, and present the limitations of the current models. Finally, the conclusion offers a brief final explanation of the observed results.

\section{Related Works}

\subsection{Classic Contrastive Classification Methods:} 
\subsubsection{SimCLR \cite{simclr}:} This method does not require additional labeled data. It only applies two different data augmentation methods to pictures, where the same image with different augmentations produces similar results while remaining mutually exclusive with other results to learn data representation. Models using SimCLR typically benefit more from increases in model depth and width, and as complexity increases, the gap with supervised methods decreases. Additionally, increasing batch size and epochs can continuously improve performance, because larger batches and prolonged training time provide more negative samples, promoting model convergence. 

\subsubsection{MoCo~v2 \cite{moco,mocov2}:} This method was originally proposed to train encoders to encode images as query vectors and key vectors. For matching image pairs, the query vector and key vector become more similar, while non-matching pairs increase their dissimilarity. It incorporates Momentum Contrast, using a queue to replace the memory bank; when the queue is full, the batch-encoded keys obtained from the newest batch data replace the oldest batch-encoded keys, making negative pairs independent of batch size but dependent on queue size. Additionally, the version 2 of MoCo incorporates an MLP projection head. 

\subsubsection{BYOL:} Bootstrap Your Own Latent (BYOL) \cite{byol} learns representations \emph{without} negative pairs, distinguishing it from the
contrastive methods above. An online network (encoder, projector, predictor) is trained to predict the projection produced by a target network (encoder, projector) whose weights are an exponential moving average (EMA) of the online network. A stop-gradient on the target, together with the predictor and the
slowly updated EMA target, lets BYOL avoid representational collapse without negatives. We adopt the PyTorch implementation available online \footnote{https://github.com/lucidrains/byol-pytorch/}.

In this project, we take all three architectures with a ResNet-18 base model in order to perform our experiments and compare their performance between their classical and hybrid counterparts. The hybrid version is defined by injecting a layer of a quantum circuit in its representation network layer.

\subsection{Hybrid Classic-Quantum Models}

\subsubsection{Quantum Self-supervised Learning (QSSL)\cite{qssl}:} Combines SimCLR with a Quantum Neural Network (QNN) by adding a parametrizable quantum circuit to the last layer of the encoded network. QSSL is classified as a Hybrid Quantum-Classical Algorithm (HQC) \cite{qssl}. HQCs consist of three elements: First, selecting a cost function based on the problem; second, selecting a quantum circuit (ansatz) based on the problem. The cost function parameters serve as parameters (variables) for the quantum gates in the ansatz, and finally a classical optimizer maximizes or minimizes the cost function\cite{vqa,vqa2}. 

\subsubsection{QuFeX\cite{qufex}:} Based on two important implementations of quantum circuits with convolutional neural networks, QCNN \cite{qcnn} and QuanNN\cite{quann}. It comprises QCNN’s circuit design, combining convolution and pooling using parameterized units, and QuanNN’s data handling mechanisms, but instead of scanning small local patches one at a time, it lets a single quantum circuit process a mix of multiple feature maps in parallel. In addition, QuFeX preserves the qubits’ output as part of the feature map. Qu-Net is a standard U-Net where its bottleneck layer is replaced by a QuFeX quantum layer and classical residual connections are added around the quantum module.  

We take both circuit designs and apply them to the classical models selected in order to compare the performance of different architectures when evaluated against the same task, in this case, fingerprint recognition.

\subsection{Fingerprint Image Preprocessing}
\subsubsection{Gabor Filter \cite{gabor}:} The Fourier transform is a powerful tool in signal processing that can help us convert images from the spatial domain to the frequency domain and extract features that are difficult to extract in the spatial domain. However, after the Fourier transform, frequency features at different positions in the image often mix together, but Gabor filters can extract local spatial frequency features and are effective texture detection tools. Based on the directional and frequency characteristics of sine waves, texture details of different directions and sizes can be obtained.
\subsubsection{Canny Edge Detection \cite{canny}:} Canny Edge Detection is a preprocessing method for extracting photo edges and reducing noise, helping extract fingerprint edge contours, making these edge features easier for models to recognize while preserving important shapes and structures.
We perform the data preprocessing and enhancement using the previously described methods.

\section{Method}
\subsection{Problem Formulation}
Let $\mathcal{D}=\{x_i\}_{i=1}^{N}$ be a set of unlabeled fingerprint images from $C=600$ identities. Self-supervised pre-training learns an encoder $f_\theta$ that maps an image to a representation $y=f_\theta(x)\in\mathbb{R}^{w}$ without using identity labels, by optimizing an SSL objective over augmented views. The frozen representation is then evaluated on \emph{identification}: for a query image $x_q$ with representation $y_q$, a $k$-nearest-neighbor classifier over a labeled gallery predicts the identity, and we report Top-1 and Top-5 accuracy over the held-out test identities. Our question is whether replacing the classical representation network with a parameterized quantum circuit $Q$ of the same width $w$ (so that $w$ equals the qubit count) improves identification, and whether any such improvement is \emph{consistent} in direction across contrastive and non-contrastive SSL objectives. Within each classical/hybrid pair, we hold the backbone, $w$, the data split, and all training hyperparameters fixed, so the representation network is the only component that changes.

\subsection{Encoder and hybrid representation network}
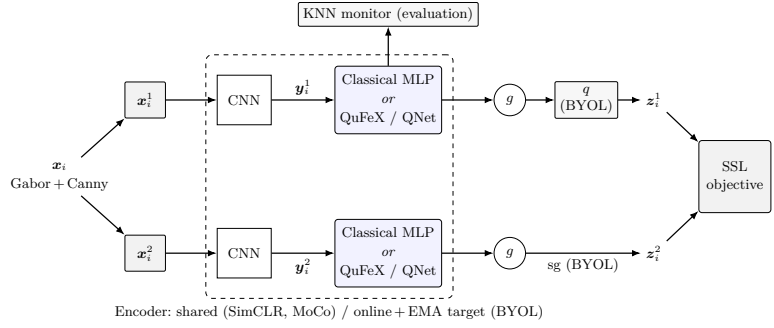
\begin{figure}[htbp]
\centering
\resizebox{0.83\textwidth}{!}{%
\begin{tikzpicture}[
  font=\small,
  io/.style={align=center},
  view/.style={draw, rounded corners=1pt, fill=gray!10, minimum width=9mm, minimum height=8mm, align=center},
  cnn/.style={draw, fill=white, minimum width=12mm, minimum height=11mm},
  rep/.style={draw, rounded corners=2pt, fill=blue!5, minimum width=22mm, minimum height=15mm, align=center},
  proj/.style={draw, circle, fill=white, minimum size=7mm, inner sep=0pt},
  small/.style={draw, rounded corners=1pt, fill=gray!6, align=center, font=\footnotesize, inner sep=3pt},
  loss/.style={draw, rounded corners=2pt, fill=gray!10, align=center, minimum width=16mm, minimum height=16mm},
  arr/.style={-{Latex[length=2mm]}, thick},
]
\node[io] (xi) at (0,0) {$\vec{x}_i$\\[1pt]\footnotesize Gabor\,+\,Canny};
\node[view] (x1) at (1.9,1.7) {$\vec{x}^{1}_i$};
\node[view] (x2) at (1.9,-1.7) {$\vec{x}^{2}_i$};
\draw[arr] (xi) -- (x1);
\draw[arr] (xi) -- (x2);
\node[cnn] (cnn1) at (4.1,1.7) {CNN};
\node[rep] (rep1) at (7.3,1.7) {Classical MLP\\[1pt]\textit{or}\\[1pt] QuFeX / QNet};
\node[proj] (p1) at (10.0,1.7) {$g$};
\node[small] (q1) at (11.7,1.7) {$q$\\(BYOL)};
\node[io] (z1) at (13.2,1.7) {$\vec{z}^{1}_i$};
\draw[arr] (x1) -- (cnn1);
\draw[arr] (cnn1) -- node[above]{$\vec{y}^{1}_i$} (rep1);
\draw[arr] (rep1) -- (p1);
\draw[arr] (p1) -- (q1);
\draw[arr] (q1) -- (z1);
\node[cnn] (cnn2) at (4.1,-1.7) {CNN};
\node[rep] (rep2) at (7.3,-1.7) {Classical MLP\\[1pt]\textit{or}\\[1pt] QuFeX / QNet};
\node[proj] (p2) at (10.0,-1.7) {$g$};
\node[io] (z2) at (13.2,-1.7) {$\vec{z}^{2}_i$};
\draw[arr] (x2) -- (cnn2);
\draw[arr] (cnn2) -- node[below]{$\vec{y}^{2}_i$} (rep2);
\draw[arr] (rep2) -- (p2);
\draw[arr] (p2) -- node[below,font=\footnotesize]{sg (BYOL)} (z2);
\begin{scope}[on background layer]
\node[draw, dashed, rounded corners, fit=(cnn1)(rep1)(cnn2)(rep2), inner sep=7pt] (enc) {};
\end{scope}
\node[font=\footnotesize] at (enc.south) [below=1pt] {Encoder: shared (SimCLR, MoCo) / online\,+\,EMA target (BYOL)};
\node[loss] (loss) at (15.0,0) {SSL\\ objective};
\draw[arr] (z1) -- (loss);
\draw[arr] (z2) -- (loss);
\node[small] (knn) at (7.3,3.6) {KNN monitor (evaluation)};
\draw[arr] (rep1.north) -- (knn);
\end{tikzpicture}%
}
\caption{Overall architecture. Each preprocessed fingerprint $\vec{x}_i$ (Gabor + Canny) is augmented into two views $\vec{x}^{1}_i, \vec{x}^{2}_i$ and passed through a shared encoder: a CNN (ResNet-18) backbone followed by a representation network that is \emph{either} a classical MLP \emph{or} the quantum module (QuFeX / QNet). A projection head $g$ maps representations to $\vec{z}^{1}_i, \vec{z}^{2}_i$ for the self supervised objective. For SimCLR and MoCo~v2, the objective is contrastive with weight sharing; In BYOL, the target branch is an EMA target, the online branch adds a predictor $q$, and a stop gradient ($sg$) is applied to the EMA target. The KNN monitor evaluates the representation $\vec{y}^{1}_i$ before the projection head.}
\vspace{-10pt}
\label{fig:arch}
\end{figure}

Figure~\ref{fig:arch} shows the overall architecture. All models share a ResNet-18 backbone adapted for single-channel input (the first convolution takes one channel), producing a $512$-dimensional feature. A linear layer compresses this to a representation of width $w$; we fix $w=8$ to match the qubit count of the hybrid models. In the classical model, the
representation network is a stack of linear layers with LeakyReLU; in the hybrid model, it is the QuFeX quantum module or QNet circuit, depending on the experiment. 
We did not search over classical alternatives (deeper/wider MLPs, or an MLP with a matching residual connection) that might close the gap without a quantum circuit; this capacity-matched ablation is left to future work.
A projection head maps the representation for the SSL loss. \emph{For all KNN evaluations, we use the encoder features before the projection head, never the projector output.}

For the QNet implementation, the quantum circuit is based on a hardware-efficient ansatz (HEA) \cite{hea}, specifically "Circuit 14" \cite{expressibility}. A HEA is a type of quantum circuit that minimizes the number of two-qubit entangling gates, respects fixed hardware connectivity, and is highly practical, as it is not designed to be problem-specific. In "Circuit 14"'s topology, each qubit connects to its nearest neighbor and the qubit three positions away, forming a circulant graph rather than a ring. This topology creates more entanglement between qubits without it being too expensive; it is trainable in strength due to its usage of parameterized gates, and allows the entangling gates to be reoriented. For the QuFex circuit, it was built based on a QCNN-structured circuit without qubit discarding.
Circuits are classically simulated, not run on real hardware; noise and decoherence are not modeled.

\subsection{Comparison of Backbone Circuits: Expressibility, Entanglement, and Residual Design}

QNet~\cite{qssl} uses "Circuit 14" from \cite{expressibility} and it is designed to be a generic and flexible building block, has a high entangling capability, and is close to an unstructured MLP, expressing no inductive bias towards any particular data structure. QuFex~\cite{qufex}, on the other hand, is inspired by QNN~\cite{qcnn}, with an explicit convolution-then-pooling hierarchy, following a classical CNN's sample efficiency relative to MLPs. 

QNet’s CRX is a parameterized two-qubit gate; the entangling strength between each pair is itself a trainable parameter, continuously interpolating from no entanglement to maximal. While QuFex’s CNOT/CZ gates are fixed, non-parameterized, maximally entangling operations. Only single-qubit rotations riding on top of them are trainable.

QNet has high expressibility and therefore can be plateau-prone, as explained in \cite{barren,connect}; however, QCNN architectures may avoid the exponential vanishing-gradient problem \cite{qcnnbp}; therefore, QuFex is structurally restricted and more plateau-resistant in theory for being QCNN-inspired, although not conclusive.

QNet addresses the issue of classic quantum circuits' linearity in pure unitary evolutions by adding a mid-circuit projective measurement; collapsing part of the state between layers is a way of injecting a non-unitary, non-linear operation into a linear circuit. QuFex is fully unitary until the final measurement and is strictly a linear quantum feature map, with all of its practical nonlinearity coming from the classical residual add and the encoding step.

QuFex's residual output composition, based on \cite{residual} skip connections, makes it easier to learn a perturbative refinement instead of a full transformation, and smooth the loss \cite{visualizing}: 
\begin{equation}
y = Q(x) + x
\end{equation}
In hybrid quantum-classical settings, the residual approach gives the network a trainable fallback to the identity map when the quantum component is poorly optimized; the classical signal is never really lost and gated behind the quantum performance. 
This also means that QuFeX's advantage over the classical baseline, where observed, cannot be cleanly separated from the benefit of the residual connection itself: a classical representation network with the same residual structure but no quantum circuit is a natural control that we do not test.
On the other hand, QNet's: 
\begin{equation}
x = f(x)
\end{equation}
has no fallback; the entire downstream representation is sent to the quantum channel, so any issues in the quantum layer can directly degrade the representation with nothing to fall back on.

\section{Experimental Settings}
\subsection{Setup}
\label{sec:setup}
All models, including SimCLR, MoCo~v2, and BYOL, use ResNet-18, batch size $128$, width $8$, and the Adam optimizer (learning rate $10^{-3}$, weight decay $10^{-6}$). BYOL is trained for $200$ epochs and MoCo~v2 for $50$ epochs; the SimCLR figures are from a separately trained run under the same backbone and width. Each configuration is a single run. As a control, we repeat the hybrid/classical comparison on CIFAR-10 ($10$ classes; random baseline Top-5 $50\%$). All models were run using a CPU Intel Xeon W5-3435X 3.10GHz, RAM Crucial ECC REG DDR5 64G, GPU NVIDIA RTX PRO 6000 96GB.

\subsection{Dataset and preprocessing}
We use the SOCOFing fingerprint dataset \cite{sokoto}: $6{,}000$ grayscale images from $600$ subjects ($10$ prints each), expanded through data augmentation. After removing excessively distorted samples, $29{,}514$ images
remain, which undergo two-stage preprocessing with a Gabor filter \cite{gabor} \footnote{https://github.com/Utkarsh-Deshmukh/Fingerprint-Enhancement-Python} and Canny edge detection\cite{canny}. The Gabor filter extracts local spatial-frequency and directional texture; Canny edge detection extracts ridge contours while reducing noise, preserving the shapes most useful for recognition. The data are split $8{:}1{:}1$ into training, validation, and test sets over the $600$ identities.

\subsection{Integration into SSL frameworks and evaluation}

The same hybrid representation network is inserted into each framework. SimCLR and MoCo~v2 use a contrastive loss over positive and negative pairs; MoCo~v2 maintains a momentum encoder and a queue of negative keys. BYOL uses no negatives, with an
EMA target ($m=0.996$), a $2$ layer projector (dimension $128$, hidden $512$), and a $2$ layer predictor. 
Here, $m$ denotes the momentum coefficient governing the EMA update of the target network's weights, $\xi \leftarrow m \xi + (1-m)\theta$, where $\theta$ are the online network's parameters.
Recognition is measured by a KNN monitor over encoder
features, reporting Top-1 and Top-5. For $600$ classes the random baseline is Top-5 $\approx 0.83\%$.

For a fair quantum versus classical comparison, we fix the width $w=8$ ($=8$ qubits), the backbone, the dataset split, the image size, and the batch size within each hybrid/classical pair. Classical width-$16/32/64$ runs are diagnostic only and are not part of the fair comparison.

\section{Results}\label{sec:results}
\subsection{Classical and Hybrid Self-Supervised Models Comparison}
For the first experiment, we compared the performance of the classical and hybrid (QuFex) versions of the three tested models, SimCLR, MoCo~v2, and BYOL, using a training dataset of approximately $30,000$ images and a testing set of $3900$ images. SimCLR and MoCo~v2 were trained for 50 epochs and BYOL for $200$ epochs. The addition of the quantum layer is expected to show an increase in performance compared to its classical counterpart.

\textbf{Table~\ref{main}} reports test KNN accuracy. In every QuFeX setting, the hybrid scores higher than its classical counterpart on Top-5. On fingerprints, SimCLR Top-5 rises from $48.43\%$ to $57.06\%$; MoCo~v2 improves Top-5 from $28.22\%$ to $40.30\%$; and BYOL improves Top-5 from $23.70\%$ to $25.55\%$. 
The same direction holds on CIFAR-10 (Top-5 $97.03\%\!\rightarrow\!97.52\%$). 
The improvement is thus 
consistent in sign for QuFeX across objectives and datasets (QNet, below, is not), although its magnitude varies and is smallest for BYOL. 
Given single runs, smaller margins may reflect run-to-run noise.
The hybrid models cost $1.5\times$ to $6\times$ more per epoch, so gains are not at equal compute. 
BYOL's lower absolute accuracy may reflect its lack of explicit negative pairs in a fine-grained, 600-identity task~\cite{wang2020}; since hyperparameters were not matched across objectives, we compare classical and hybrid models only within each objective.

\begin{table}[!t]
\centering
\caption{Test KNN Top-5 accuracy (\%) on encoder features (before the projection
head). Hybrid uses the QuFeX quantum
module; width $=8$ ($=8$ qubits) for the fingerprint runs.Training budgets: SimCLR 100 epochs, MoCo~v2 50 epochs, BYOL
200 epochs; the CIFAR-10 control uses BYOL trained for 200 epochs. All values are
from single runs, taken from the final-epoch model. Better value in each
classical/hybrid pair in bold. Per-seed BYOL statistics are given in
Section~\ref{sec:byol-seeds} (Tables~\ref{tab:byol-seeds}--\ref{tab:byol-allseeds}).}
\label{main}
\small
\begin{tabular}{|l|l|}\hline
Method (variant) & Top-5 \\\hline
\multicolumn{2}{|l|}{\emph{SOCOFing (600 cls.; rand.\ 0.83)}}\\\hline
SimCLR, classical      & 48.43                 \\
SimCLR, hybrid         & \textbf{57.06}        \\
MoCo v2, classical     & 28.22                 \\
MoCo v2, hybrid        & \textbf{40.30}        \\
BYOL, classical        & 23.70                 \\
BYOL, hybrid           & \textbf{25.55}        \\\hline
\multicolumn{2}{|l|}{\emph{CIFAR-10 (10 cls.; rand.\ 50)}}\\\hline
BYOL, classical        & 97.03                 \\
BYOL, hybrid (QuFeX)   & \textbf{97.52}        \\\hline
\end{tabular}
\end{table}
\vspace{-5pt}

\begin{figure}[!t]
\centering
\includegraphics[width=0.6\textwidth]{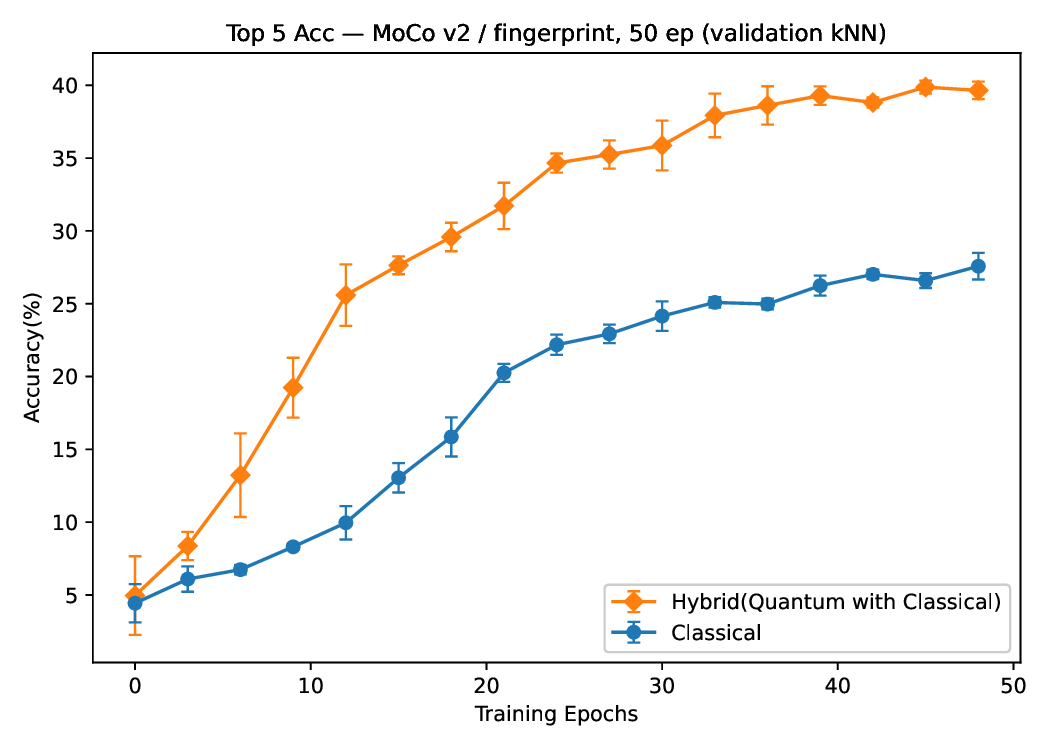}
\caption{MoCo~v2 on fingerprints (SOCOFing, 50 epochs), classical vs.\ hybrid (QuFeX) at width $8$. The hybrid separates from the classical baseline within the first ten epochs and stays above it for the whole run on validation Top-5, so the advantage is not an artefact of the single best epoch. Markers are single-run values (seed~0); error bars show $\pm1$ s.d.\ of the metric over a forward window of a few consecutive epochs, i.e.\ within-run epoch-to-epoch volatility, not confidence intervals, across-seed variance, or the significance of the hybrid--classical gap.}
\vspace{-1pt}
\label{fig:moco}
\end{figure}

\begin{figure}[!t]
\centering
\includegraphics[width=0.6\textwidth]{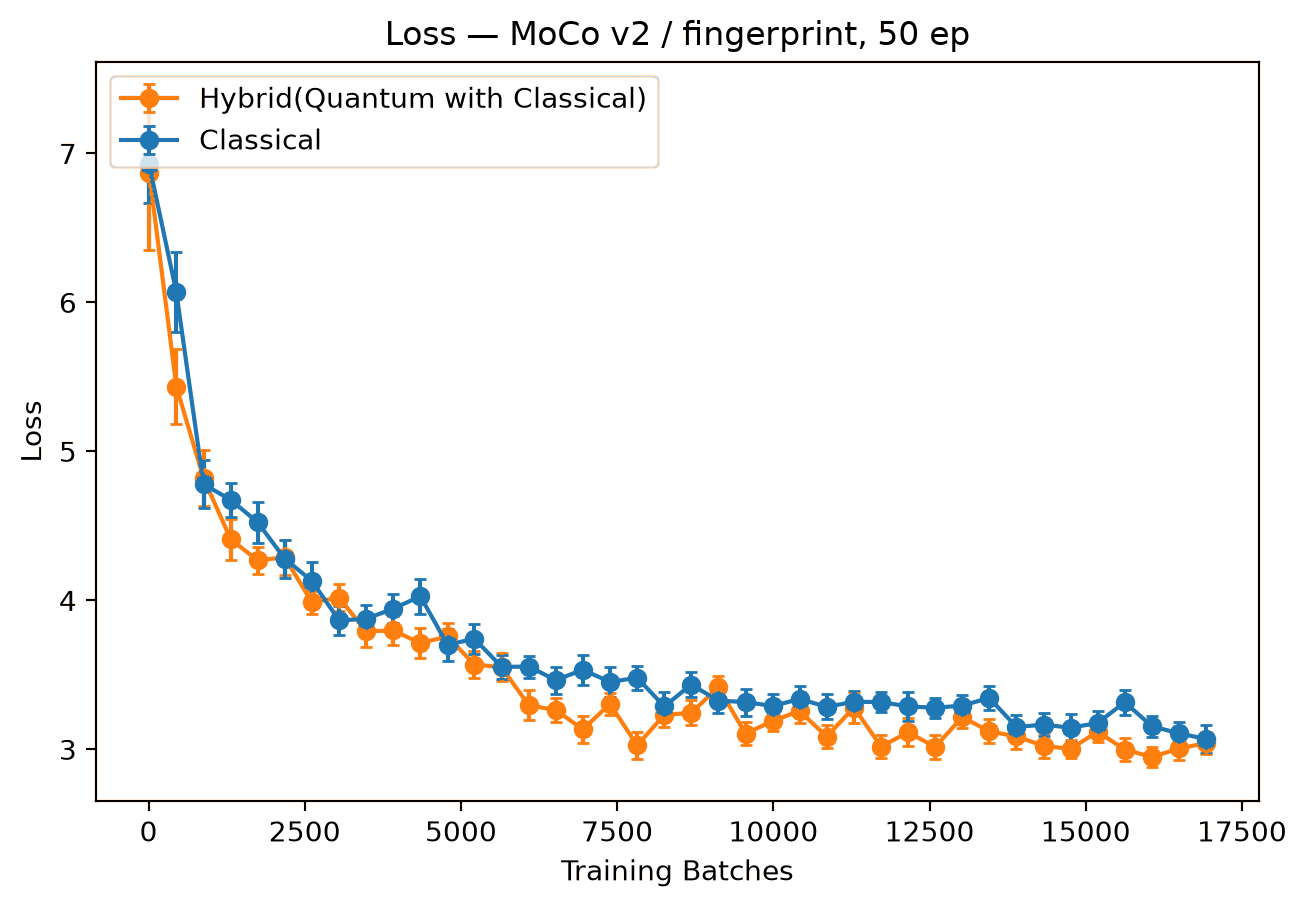}
\caption{MoCo~v2 training loss (InfoNCE) on fingerprints, classical vs.\ hybrid (QuFeX). The hybrid loss is consistently lower, so loss and recognition (Fig.~\ref{fig:moco}) agree. Markers are single-run values (seed~0); error bars show $\pm1$ s.d.\ of the loss over a forward window of ${\approx}20$ consecutive batches, i.e.\ within-run batch-to-batch volatility inflated by the training-curve slope, not confidence intervals or across-seed variance.}
\label{fig:moco-loss}
\vspace{15pt}
\begin{minipage}[t]{0.43\textwidth}
\centering
\includegraphics[width=\textwidth]{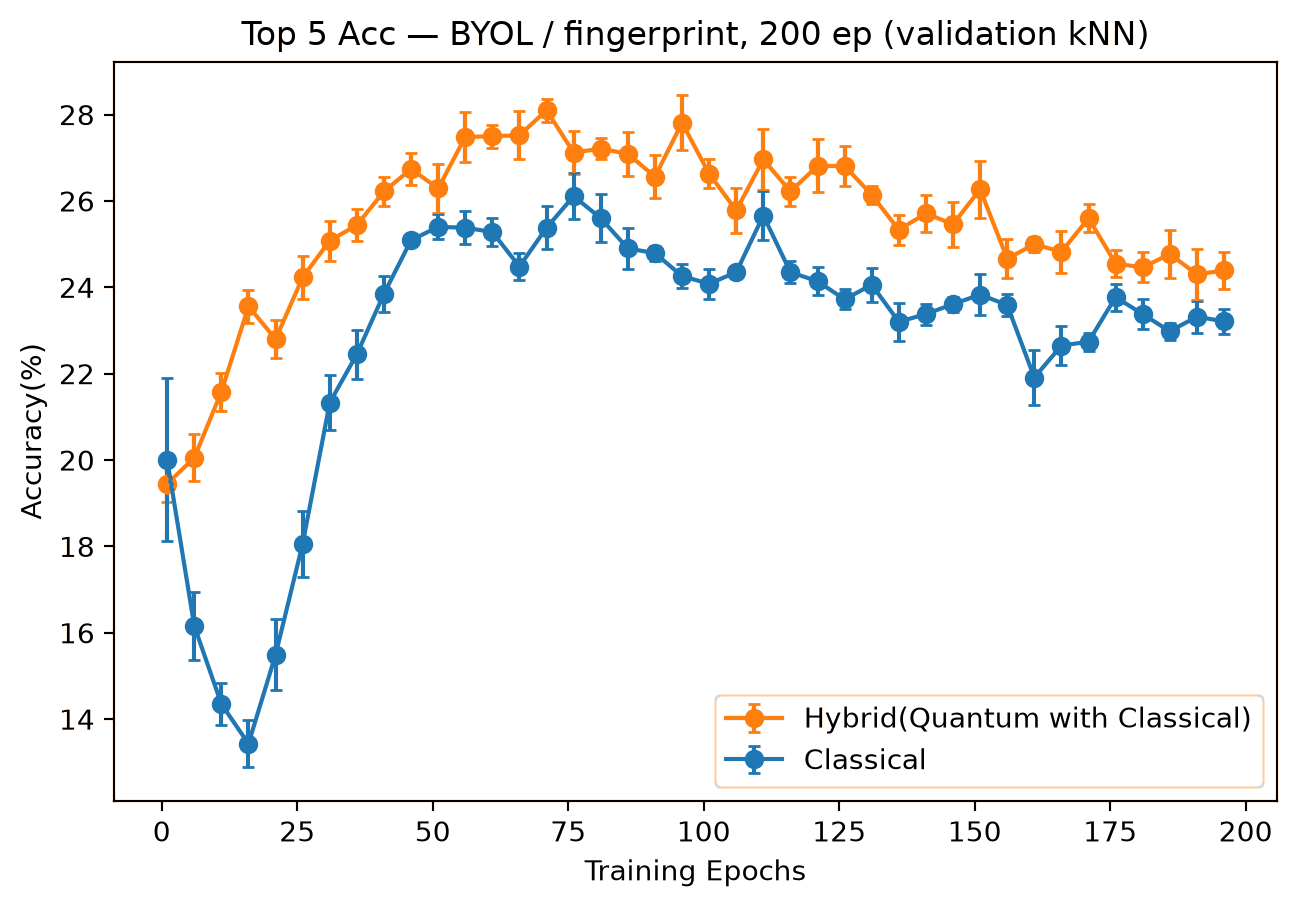}
\vskip 4pt 
Fingerprint (SOCOFing), 200 epochs \label{fig:byol-fp}
\end{minipage}
\hfil
\begin{minipage}[t]{0.43\textwidth}
\centering
\includegraphics[width=\textwidth]{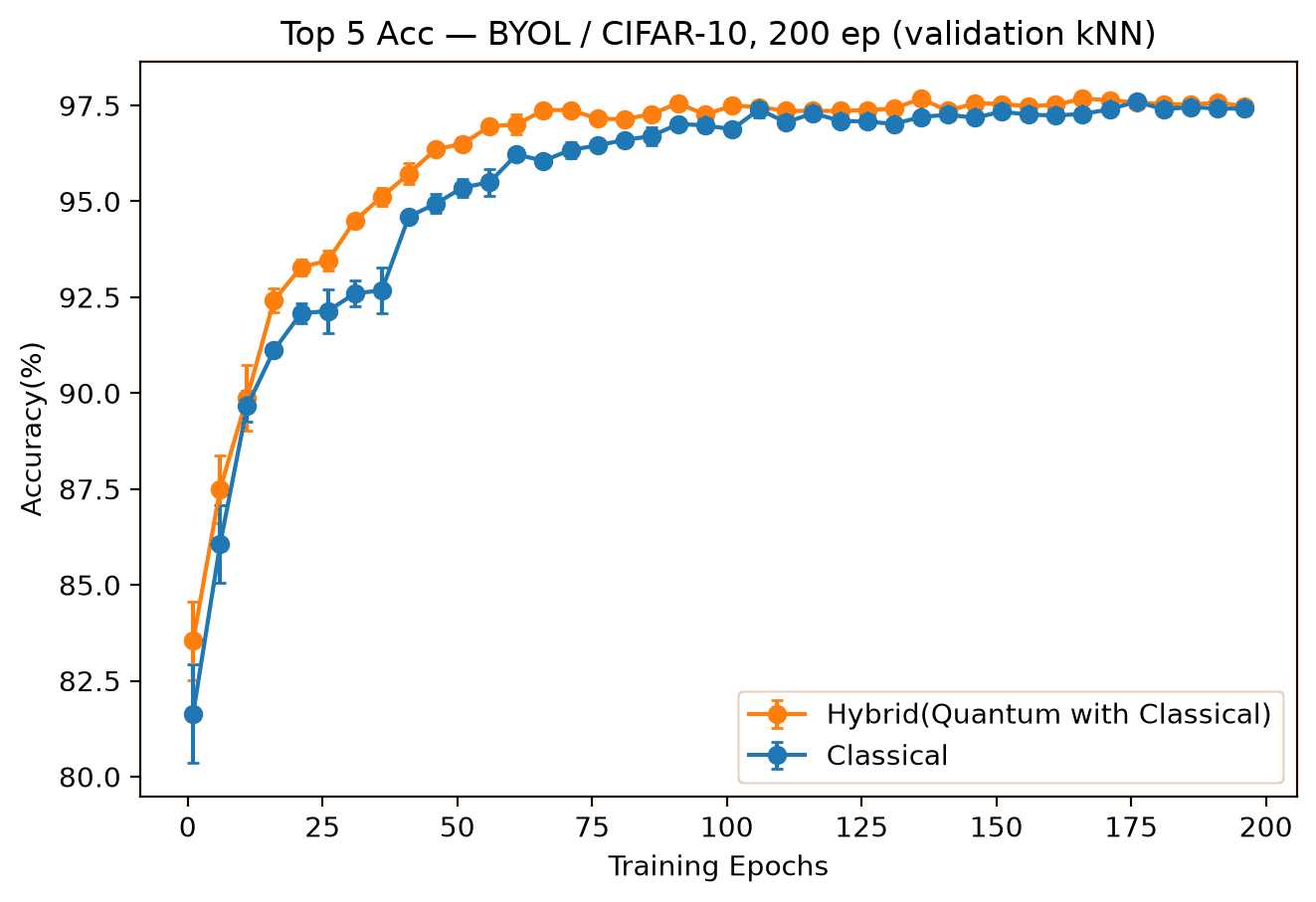}
\vskip 4pt 
CIFAR-10 control, 200 epochs \label{fig:byol-cifar}
\end{minipage}
\caption{BYOL validation Top-5 kNN, classical vs.\ hybrid (QuFeX). (a)~On fingerprints, the hybrid rises smoothly while the classical model dips to about $13\%$ near epoch~15 before recovering. (b)~On the CIFAR-10 control both saturate near $97.5\%$, but the hybrid reaches the plateau earlier. Markers are single-run values (seed~0); error bars show $\pm1$ s.d.\ over a forward window of a few consecutive epochs (within-run volatility), not confidence intervals, across-seed variance, or gap significance.}
\vspace{-10pt}
\label{fig:byol}
\end{figure}

\paragraph{Training dynamics.}
Figures~\ref{fig:moco}--\ref{fig:byol} show the per-epoch behaviour behind the final numbers in Table~\ref{main}, which is
evidence that the gap is not a single-epoch artefact, though still from one seed.
For MoCo~v2 (Fig.~\ref{fig:moco}), the hybrid model separates from the classical baseline within the first ten epochs and stays above it for the entire run on validation Top-5; it also converges faster, peaking near epoch~20 while the classical model is still improving. Its InfoNCE training loss (Fig.~\ref{fig:moco-loss}) is correspondingly lower, so the training objective and the recognition metrics agree. For BYOL (Fig.~\ref{fig:byol}\,(a)) the hybrid rises smoothly and monotonically on fingerprints, whereas the classical model dips to about $13\%$ Top-5 near epoch~15 before recovering, so the quantum module also yields more stable training, not only higher accuracy. 
On the CIFAR-10 control (Fig.~\ref{fig:byol}\,(b)) both models saturate near $97.5\%$ Top-5, but the hybrid reaches that plateau earlier. Across a contrastive method (MoCo~v2) and a non-contrastive one (BYOL), and across two datasets, the hybrid curve lies above its classical counterpart for essentially all of training,
reinforcing Table ~\ref{main}.'s sign, though not its
significance.

\subsection{Classical and Hybrid SimCLR and MoCo~v2 with different epoch lengths}
For the second experiment, we compared the performance of the classical and hybrid (QuFex) versions of SimCLR when running at different epoch lengths. Additionally, we tested the performance of SimCLR with one layer using QNet as the quantum circuit; however, in this instance, it was only tested for 15 epochs due to time and hardware constraints. 
This held for QuFeX but not QNet.
Results are presented in the table ~\ref{tab2}, and figures~\ref{fig1}  and ~\ref{fig2} 

We additionally report two diagnostic experiments in Appendix~A: an upstream depth sweep and a downstream linear-versus-KNN comparison for the QNet-based SimCLR. They motivate the width-matched QuFeX study here, since they show that a quantum representation network helps only at depth $\ge 2$ and scores lower the classical baseline at a single layer (Appendix Table~\ref{tab3}); this is why we adopt QuFeX for the single-layer injection used throughout the main experiments.

\textbf{Table~\ref{tab2}} shows the highest percentage obtained with each model definition after running 15, 50, 100, and 200 epochs. Due to time and hardware constraints, the SimCLR-Classic model with 200 epochs, as well as the original version of QSSL (using QNet as the quantum layer) was only run for 15 epochs, so its converged performance is unknown.

In the QuFeX instances, the SimCLR-QuFex model consistently scores higher than the other model variations, with the highest result obtained at $57\%$. QNet scores lower than classical SimCLR at its only tested budget (Table~\ref{tab2}). The same pattern can be found in the MoCo~v2 experiments (Table~\ref{tab5} where the QuFex variant scores higher than the classical variant, with the highest result obtained at approximately $52\%$. However, it is also observable that the advantage presented by the MoCo-QuFex model for the larger number of epochs against its classical counterpart is not as pronounced as the smaller number of epochs.

\begin{table}[!t]
\centering
\caption{Highest Percentage of Top-5 Test Accuracy in SimCLR classical and hybrid models. In bold, the highest result obtained from all the runs.}\label{tab2}
\begin{tabular}{|l|l|l|l|l|}\hline
Model&  15 Epochs& 50 Epochs& 100 Epochs&200 Epochs\\
\hline
SimCLR-Classic&  34.98& 44.25& 48.43&51.23\\\hline
SimCLR-QuFex&  45.49& 55.25& \textbf{57.06}&56.13\\\hline
SimCLR-QNet& 32.27& --- & --- & ---\\\hline
\end{tabular}
\end{table}

\begin{table}[!t]
\centering
\caption{Highest Percentage of Top-5 Test Accuracy in MoCo~v2 classical and hybrid models. In bold, the highest result obtained from all the runs.}\label{tab5}
\begin{tabular}{|l|l|l|l|l|}\hline
Model&  15 Epochs& 50 Epochs& 100 Epochs&200 Epochs\\
\hline
MoCo-Classic&  21.28& 28.22& 41.91&47.77\\\hline
MoCo-QuFex&  30.52& 40.30& 45.03&\textbf{51.79}\\\hline
\end{tabular}
\end{table}

\textbf{Tables~\ref{tab6}, ~\ref{tab7}, ~\ref{tab8} and ~\ref{tab9}} denote the loss and accuracy values obtained at specific epoch numbers during training for both contrastive models, SimCLR and MoCo v2, comparing the hybrid and classical versions of each architecture. From all the results we can observe, the QuFex version in all instances obtains the highest accuracy and lowest loss regardless of the epoch number. 

\begin{table}[!t]
\centering
\setlength{\tabcolsep}{8pt}
\caption{Comparison Table for epochs=15 denoting the testing values at specific epoch numbers}\label{tab6}
\begin{tabular}{|l|l|l|l|l|l|l|l|l|}\hline
 \multicolumn{9}{|c|}{15-Epochs}\\\hline\hline
 & \multicolumn{4}{|c|}{SimCLR}& \multicolumn{4}{|c|}{MoCo}\\\hline\hline
 & \multicolumn{2}{|c|}{Classical}& \multicolumn{2}{|c|}{Hybrid}& \multicolumn{2}{|c|}{Classical}&\multicolumn{2}{|c|}{Hybrid}\\\hline\hline
 Epoch N°&  Loss&Acc&  Loss&Acc&  Loss&Acc& Loss&Acc\\\hline\hline
1 Epoch&  3.51&23.47& 3.27&24.32& 6.39&11.28&6.33&14.53\\
\hline
10 Epoch&  0.96&34.75& 0.78&44.41& 4.31&18.90&3.93&22.75\\\hline
15 Epoch&  0.82&34.55& 0.64&\textbf{45.24}& 4.02&21.27&3.68&\textbf{30.52}\\\hline
\end{tabular}
\end{table}

\begin{table}[!t]
\centering
\setlength{\tabcolsep}{8pt}
\caption{Comparison Table for epochs=50 denoting the testing values at specific epoch numbers}\label{tab7}
\begin{tabular}{|l|l|l|l|l|l|l|l|l|}\hline
 \multicolumn{9}{|c|}{50-Epochs}\\\hline\hline
 & \multicolumn{4}{|c|}{SimCLR}& \multicolumn{4}{|c|}{MoCo}\\\hline\hline
 & \multicolumn{2}{|c|}{Classical}& \multicolumn{2}{|c|}{Hybrid}& \multicolumn{2}{|c|}{Classical}&\multicolumn{2}{|c|}{Hybrid}\\\hline\hline
 Epoch N°&  Loss&Acc&  Loss&Acc&  Loss&Acc& Loss&Acc\\\hline\hline
1 Epoch&  3.56&16.65& 3.17&20.48& 6.70&3.00&6.41&2.27\\
\hline
30 Epoch&  0.59&42.89& 0.50&51.61& 3.26&23.65&2.78&34.29\\\hline
50 Epoch&  0.48&43.71& 0.43&\textbf{51.85}& 2.90&28.22&2.47&\textbf{40.07}\\\hline
\end{tabular}
\end{table}

\begin{table}[!t]
\centering
\setlength{\tabcolsep}{8pt}
\caption{Comparison Table for epochs=100 denoting the testing values at specific epoch numbers}\label{tab8}
\begin{tabular}{|l|l|l|l|l|l|l|l|l|}\hline
 \multicolumn{9}{|c|}{100-Epochs}\\\hline\hline
 & \multicolumn{4}{|c|}{SimCLR}& \multicolumn{4}{|c|}{MoCo}\\\hline\hline
 & \multicolumn{2}{|c|}{Classical}& \multicolumn{2}{|c|}{Hybrid}& \multicolumn{2}{|c|}{Classical}&\multicolumn{2}{|c|}{Hybrid}\\\hline\hline
 Epoch N°&  Loss&Acc&  Loss&Acc&  Loss&Acc& Loss&Acc\\\hline\hline
1 Epoch&  3.69&16.11& 3.23&25.33& 7.43&2.71&6.35&7.59\\
\hline
30 Epoch&  0.61&41.11& 0.49&49.19& 3.09&31.52&2.67&39.30\\\hline
50 Epoch&  0.49&45.00& 0.41&53.40& 2.66&41.23&2.47&43.12\\\hline
 70 Epoch& 0.40& 44.67& 0.36& 55.51& 2.47& 41.60& 2.34&\textbf{45.03}\\\hline
 100 Epoch& 0.36& 46.73& 0.33& \textbf{56.44}& 2.35& 41.57& 2.21&43.66\\\hline
\end{tabular}
\end{table}

\begin{table}[!t]
\centering
\setlength{\tabcolsep}{8pt}
\caption{Comparison Table for epochs=200 denoting the testing values at specific epoch numbers}\label{tab9}
\begin{tabular}{|l|l|l|l|l|l|l|l|l|}\hline
 \multicolumn{9}{|c|}{200-Epochs}\\\hline\hline
 & \multicolumn{4}{|c|}{SimCLR}& \multicolumn{4}{|c|}{MoCo}\\\hline\hline
 & \multicolumn{2}{|c|}{Classical}& \multicolumn{2}{|c|}{Hybrid}& \multicolumn{2}{|c|}{Classical}&\multicolumn{2}{|c|}{Hybrid}\\\hline\hline
 Epoch N°&  Loss&Acc&  Loss&Acc&  Loss&Acc& Loss&Acc\\\hline\hline
1 Epoch&  3.66&15.34& 3.50&20.68& 6.57&3.74&6.28&22.10\\
\hline
50 Epoch&  0.56&40.72& 0.45&52.96& 2.54&38.29&2.35&47.25\\\hline
100 Epoch&  0.40&44.02& 0.37&54.04& 22.25&43.09&2.21&46.24\\\hline
 150 Epoch& 0.34& 47.61& 0.34& \textbf{55.02}& 2.15& 43.66& 2.14&\textbf{48.25}\\\hline
 200 Epoch& 0.32& 48.67& 0.33& 51.43& 2.09& 47.68& 2.10&46.94\\\hline
\end{tabular}
\end{table}
\vspace{-5pt}

The graphical results of the model's performance when injecting one layer of QuFex into a classic SimCLR and MoCo~v2 architectures are shown in Fig.~\ref{fig1} and Fig.~\ref{moco-hybrid}, where it is also observable the stable and consistent decrease in the training loss for all models. While for the classical SimCLR and MoCo~v2 models in Fig.~\ref{fig2} and  Fig.~\ref{moco-classic}, we can observe the same consistent behavior but at a slower rate, meaning it takes a larger number of epochs for it to reach the same level of accuracy as the hybrid models.
As with the other results in this section, all figures report single training curves without repeated-seed variance
bands.

\begin{figure}[!t]
\centering
\includegraphics[width=\textwidth, height=0.4\textheight, keepaspectratio]{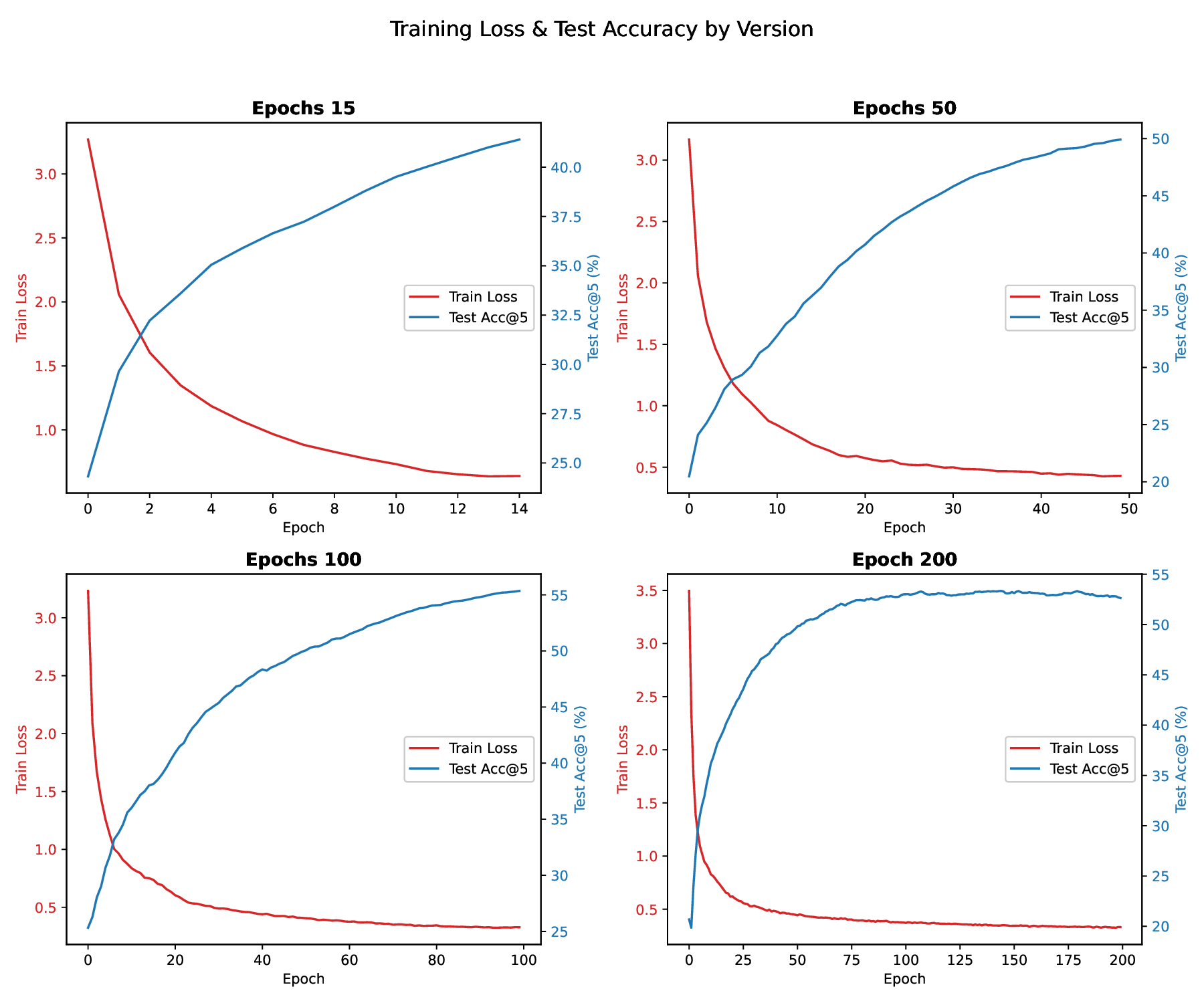}
\caption{Hybrid Quantum-Classical accuracy versus loss for SimCLR models with one QuFex layer injected at different epoch sizes.} \label{fig1}
\end{figure}
\begin{figure}[!t]
\centering
\includegraphics[width=\textwidth, height=0.4\textheight, keepaspectratio]{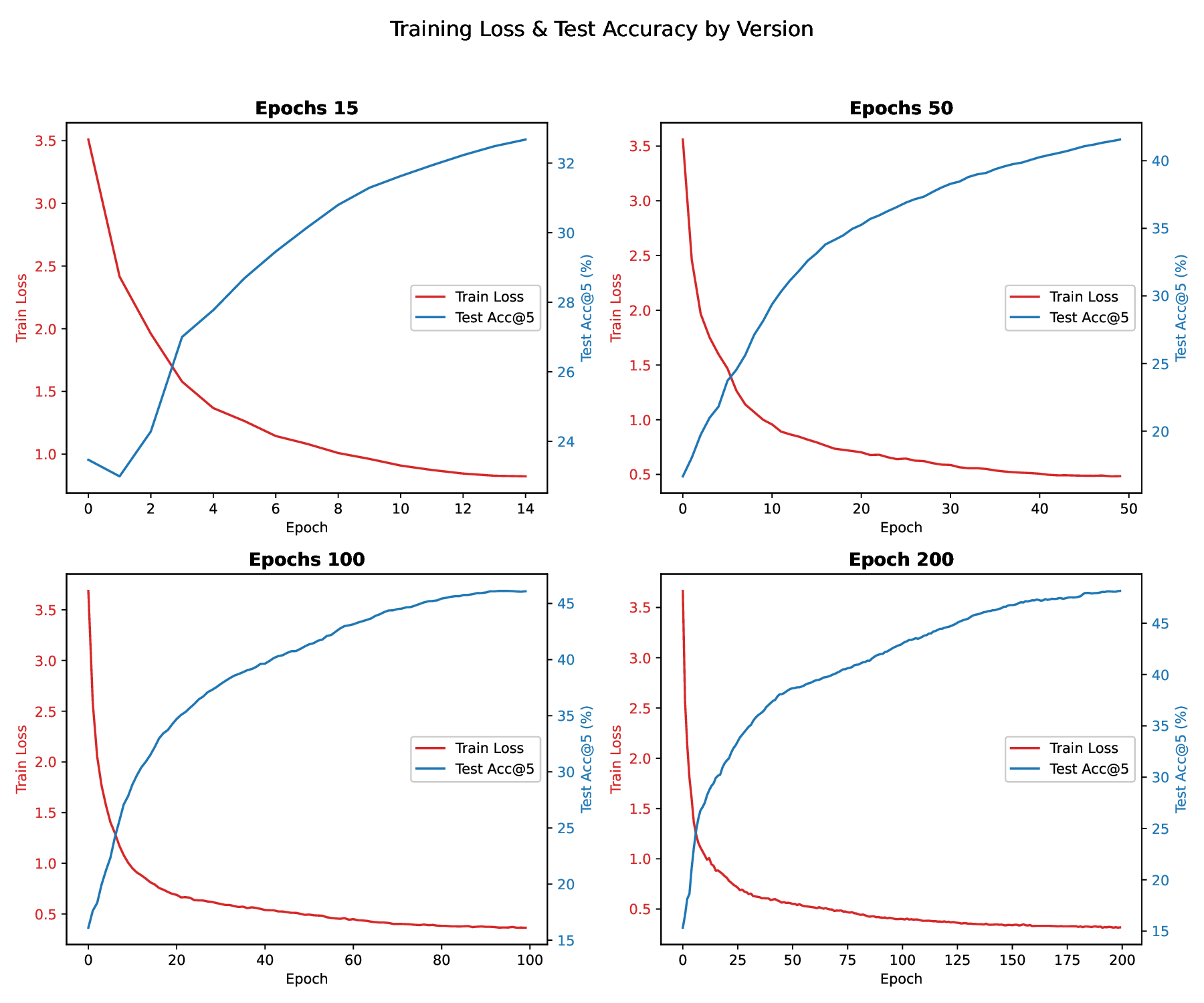}
\caption{Classical SimCLR accuracy versus loss at different epoch sizes.} \label{fig2}
\end{figure}

\begin{figure}[!t]
\centering
\includegraphics[width=\textwidth, height=0.4\textheight, keepaspectratio]{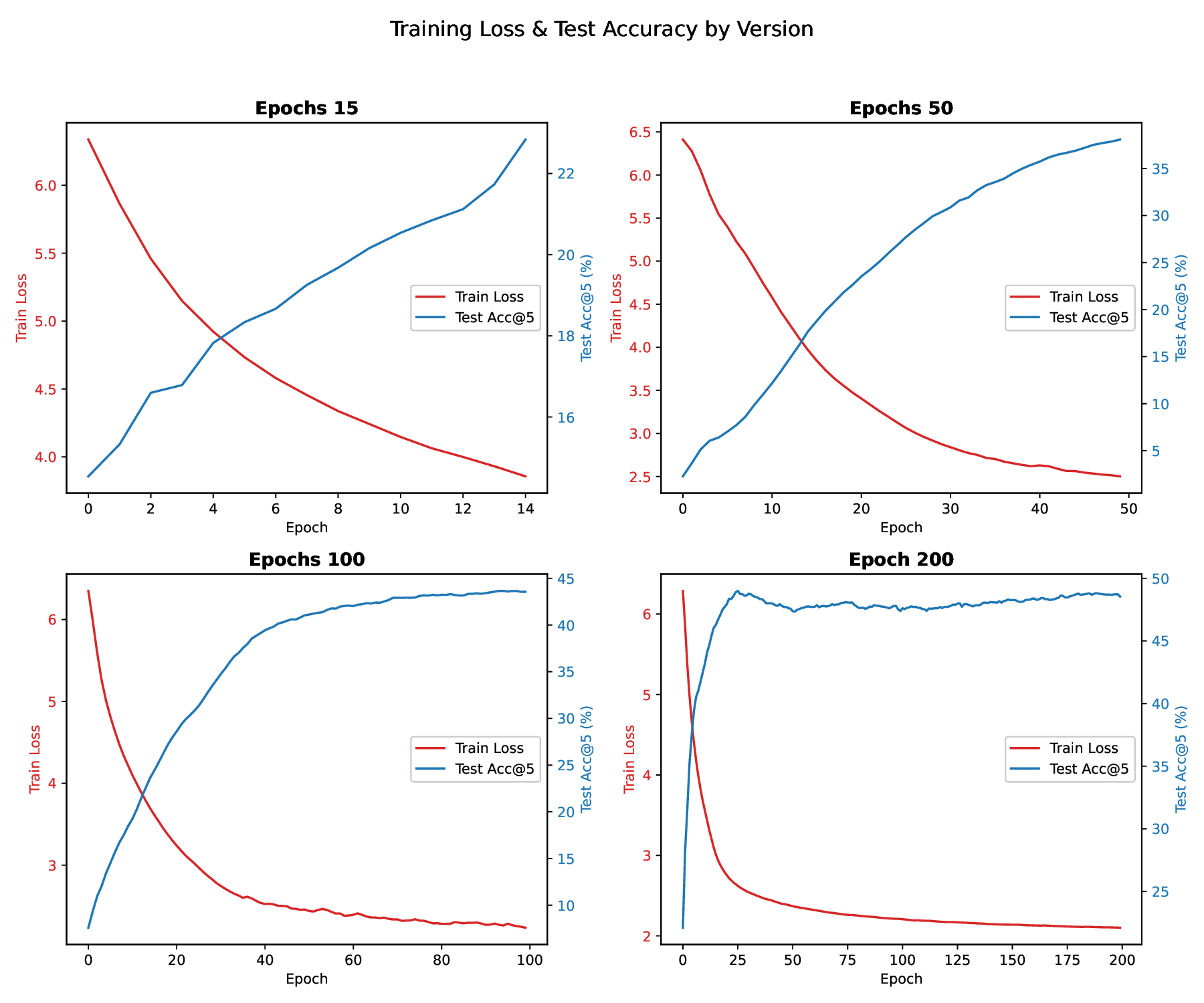}
\caption{Hybrid Quantum-Classical accuracy versus loss for MoCo~v2 models with one QuFex layer injected at different epoch sizes.} \label{moco-hybrid}
\end{figure}
\begin{figure}[!t]
\centering
\includegraphics[width=\textwidth, height=0.4\textheight, keepaspectratio]{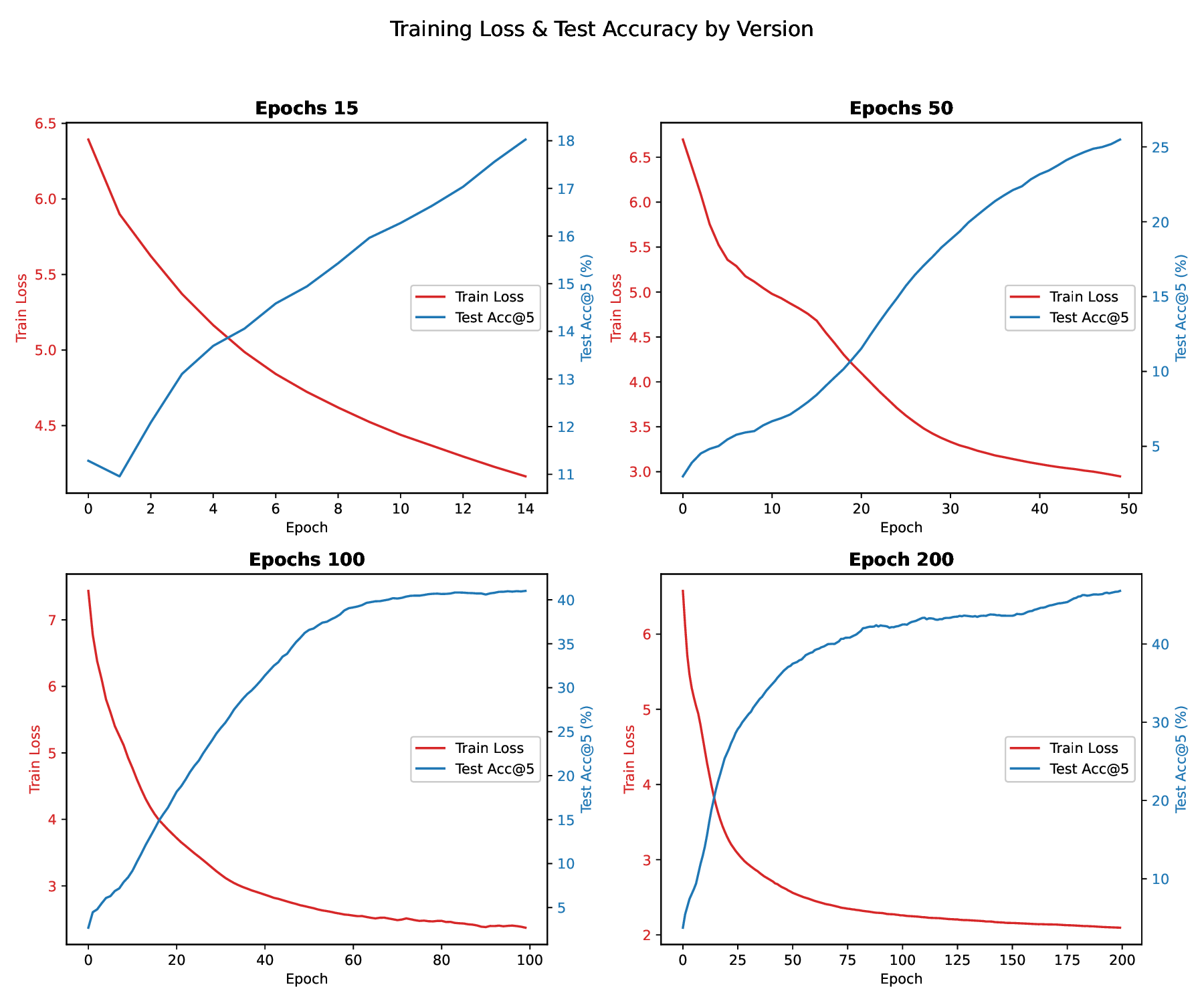}
\caption{Classical MoCo~v2 accuracy versus loss at different epoch sizes.} \label{moco-classic}
\end{figure}

\begin{figure}[!t]
\centering
\includegraphics[width=\textwidth, height=\textheight, keepaspectratio]{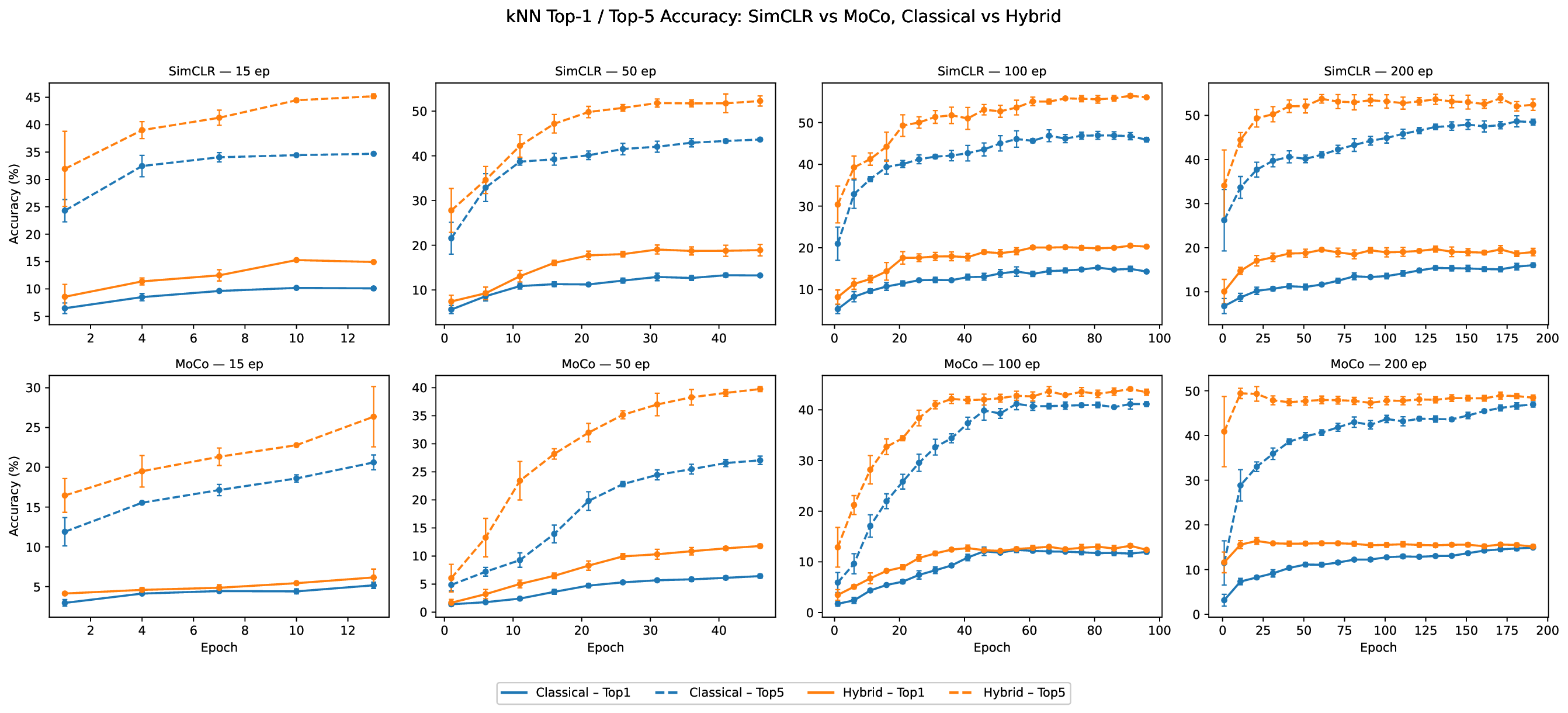}
\caption{Test Top-1/Top-5 accuracy for SimCLR and MoCo v2, classical vs. hybrid (QuFeX), across four different epoch lenghts (15, 50, 100, 200).} \label{sim-moco-acc}
\end{figure}

Additionally, we denote also obtain the kNN Top-1 accuracy results. \textbf{Figure~\ref{sim-moco-acc}} denotes the test Top-1 and Top-5 accuracies for SimCLR and MoCo v2, classical vs. hybrid (QuFeX), across four different epoch lenghts (15, 50, 100, 200). The hybrid version scores higher than its classical counterpart on both metrics, for both backbones, in all instances, the gap is present from early epochs and persists through the full run rather than appearing only at a single checkpoint.

\subsection{BYOL across multiple seeds}
\label{sec:byol-seeds}

Among the three objectives, BYOL shows the smallest single-run gap between the classical and hybrid models. In Table~\ref{main}, QuFeX improves BYOL's Top-5 accuracy by only $+1.85$ points ($23.70\%\rightarrow25.55\%$), against $+8.63$ for SimCLR and $+12.08$ for MoCo~v2. A margin this small is comparable to the run-to-run variation seen elsewhere in our experiments, so a single run cannot establish whether the quantum module helps BYOL at all. We therefore repeated the BYOL experiment with three seeds ($0,1,2$) at $15$, $50$, $100$, and $200$ epochs, for both the classical and the QuFeX model, and report the aggregate results below.

Table~\ref{tab:byol-seeds} summarizes the outcome: for each epoch budget it gives the mean $\pm$ standard deviation of the final-epoch Top-5 accuracy over the three seeds, the mean paired gap $\Delta=\text{QuFeX}-\text{Classical}$, and the number of seeds in which QuFeX wins. The benefit is \emph{budget-dependent} rather than uniform. At $50$ epochs the classical model is higher on average ($\Delta=-1.51$, QuFeX ahead in only $1/3$ seeds); at $15$ and $100$ epochs QuFeX leads on average ($\Delta=+2.75$ and $+0.66$) but the seed ranges overlap and the difference is not significant; and only at $200$ epochs does QuFeX exceed the classical baseline across all three seeds with non-overlapping ranges ($\Delta=+2.59$; paired $t=3.57$, $p\approx0.07$ at $n=3$). For BYOL, then, the quantum advantage is reliable only at the longest budget, and the single-run margin reported in Table~\ref{main} should be read as directional rather than as a reliable effect at shorter budgets.

\begin{table}[!t]
\centering
\caption{BYOL final-epoch test KNN Top-5 accuracy (\%) over three seeds (0,1,2), classical vs.\ hybrid (QuFeX), width 8. Mean $\pm$ sample std.\ ($n=3$, ddof$=1$). $\Delta$ is the mean paired gap (Q$-$C) and ``QuFeX wins'' the number of seeds in which QuFeX is higher. Bold marks the higher mean when the seed ranges do not overlap.}
\label{tab:byol-seeds}
\begin{tabular}{|c|c|c|c|c|}\hline
Epochs & Classical & QuFeX & $\Delta$ (Q$-$C) & QuFeX wins \\\hline
15  & 19.81 $\pm$ 3.44 & 22.56 $\pm$ 3.59 & $+2.75$ & 3/3 \\\hline
50  & 26.02 $\pm$ 2.90 & 24.51 $\pm$ 2.88 & $-1.51$ & 1/3 \\\hline
100 & 25.17 $\pm$ 2.46 & 25.83 $\pm$ 1.71 & $+0.66$ & 2/3 \\\hline
200 & 23.75 $\pm$ 0.12 & \textbf{26.34 $\pm$ 1.37} & $+2.59$ & 3/3 \\\hline
\end{tabular}
\end{table}

Table~\ref{tab:byol-allseeds} lists the per-seed values behind these means, with Top-1 included for completeness. The per-seed numbers expose the variability directly: seed~1 favors the classical model at $50$ and $100$ epochs---its classical Top-5 of $29.35\%$ at $50$ epochs is the highest BYOL classical result anywhere---whereas seed~2 favors QuFeX at every budget. Because this seed-to-seed spread is larger than the mean gap at every budget except $200$ epochs, no single seed is representative, which is why we report the aggregate statistics of Table~\ref{tab:byol-seeds} rather than any individual run.

\begin{table}[!ht]
\centering
\setlength{\tabcolsep}{14pt}
\renewcommand{\arraystretch}{1.5}
\caption{Complete BYOL results: final-epoch test KNN accuracy (\%) for every seed. Classical vs.\ hybrid (QuFeX), width 8, on SOCOFing. Top-1 is included for completeness, while the paper's analysis uses Top-5; aggregate mean\,$\pm$\,std is given in Table~\ref{tab:byol-seeds}.}
\label{tab:byol-allseeds}
\resizebox{\textwidth}{!}{%
\begin{tabular}{|c|c|c|c|c|c|c|c|}\hline
\multirow{2}{*}{Ep.} & \multirow{2}{*}{Model} & \multicolumn{2}{c|}{Seed 0} & \multicolumn{2}{c|}{Seed 1} & \multicolumn{2}{c|}{Seed 2}\\\cline{3-8}
 & & Top-1 & Top-5 & Top-1 & Top-5 & Top-1 & Top-5 \\\hline
\multirow{2}{*}{15}  & Classical & 6.95 & 17.35 & 11.13 & 23.74 & 7.98 & 18.33 \\\cline{2-8}
                     & QuFeX     & 8.11 & 18.45 & 10.35 & 24.10 & 10.77 & 25.11 \\\hline
\multirow{2}{*}{50}  & Classical & 10.04 & 24.01 & 13.21 & 29.35 & 11.02 & 24.70 \\\cline{2-8}
                     & QuFeX     & 9.19 & 22.62 & 9.81 & 23.09 & 11.81 & 27.83 \\\hline
\multirow{2}{*}{100} & Classical & 10.98 & 24.14 & 12.57 & 27.97 & 10.51 & 23.39 \\\cline{2-8}
                     & QuFeX     & 10.84 & 27.34 & 10.82 & 23.97 & 11.38 & 26.18 \\\hline
\multirow{2}{*}{200} & Classical & 10.69 & 23.70 & 10.46 & 23.88 & 10.19 & 23.67 \\\cline{2-8}
                     & QuFeX     & 11.04 & 25.55 & 12.38 & 27.92 & 11.05 & 25.55 \\\hline
\end{tabular}%
}
\end{table}
\FloatBarrier

\section{Discussion}
\subsection{A consistent but modest quantum benefit}
The central observation is one of  \emph{directional consistency}: inserting the QuFeX module improves KNN recognition in every setting we tested: the contrastive SimCLR and MoCo~v2, the non-contrastive BYOL, and the CIFAR-10 control 
— QNet did not show this pattern at the single-layer depth tested (Section~\ref{sec:results})
Because BYOL uses no negative pairs, this indicates that the benefit is not a property of contrastive repulsion but rather of the representation produced by the quantum module. 
QuFeX also adds a residual connection and encoding step beyond the classical baseline, so this benefit may not be purely quantum in origin.
Beyond final accuracy, the hybrid also converges faster and, for BYOL, trains more stably (Figs.~\ref{fig:moco}--\ref{fig:byol}). 
At the same time, we do not claim a large or state-of-the-art improvement: absolute fingerprint accuracy is modest, and the per-setting margins are small, from about $+0.49$ Top-5 for BYOL to about $+11$ Top-5 for SimCLR.

\subsection{Why absolute accuracy is modest, and QNet versus QuFeX}
Fingerprint recognition here is a $ 600$-class problem on small grayscale images,
compressed through an $8$ dimensional ($8$ qubit) bottleneck, with a limited
epoch budget for the quantum models; these factors bound absolute accuracy for
both classical and hybrid models. We used QuFeX rather than QNet because QuFeX is
structurally restricted and, being QCNN-inspired, is expected to be more
resistant to barren plateaus \cite{qcnnbp}, whereas the highly expressive QNet
can be plateau prone \cite{barren}; consistent with this, our depth sweep
(Appendix Table~\ref{tab3}) shows QNet helping only at depth $\ge 2$ and
scores lower than the classical baseline at the single layer we use. QuFeX's residual composition (Eq.~1) further provides a trainable fallback that QNet (Eq.~2) lacks. Completing the compute-limited QNet runs and comparing the two circuits empirically is left to future work.

 \subsection{Limitations}
First, each configuration is a single run (seed~0); the bands in Figs.~\ref{fig:moco}--\ref{fig:byol} show only within-run batch/epoch volatility, so we cannot report across-seed variance, and the smaller margins (notably BYOL Top-5) are within plausible run-to-run variance. It is the
agreement in direction across independent settings, rather than any single margin, that supports the claim. 
This is thus a claim about effect sign, not statistical significance.
Second, the hybrid adds the \emph{entire} QuFeX module (quantum circuit, encoding, and classical residual); although the representation width is matched, the module's capacity is not, so part of the gain may be attributable to added capacity rather than to the quantum circuit itself. A capacity-matched classical module is the natural ablation. 
We do not run it here. 
Third, the benefit comes at a compute cost of $1.5\times$ to $6\times$ per epoch under simulation, so it is not a compute saving.
Fourth, circuits are classically simulated, not run on real hardware. Fifth, epoch budgets differ across objectives (50, 200, 15), confounding cross-objective margin comparisons. Sixth, QNet scored lower than classical SimCLR at matched depth, so the positive claim is specific to QuFeX.

\section{Conclusion}
We presented a unified hybrid framework that inserts the same quantum
feature extraction module into three self-supervised objectives: SimCLR, MoCo~v2, and BYOL for fingerprint recognition. Under a width-matched comparison, the hybrid model consistently scored higher its classical counterpart on KNN Top-5 across all three objectives and on a CIFAR-10 control, indicating that
the benefit of the quantum module is not specific to the contrastive objective. The margins are modest and obtained from single runs at higher compute cost. Confirming them with multiple seeds and error bars, isolating the quantum contribution with capacity-matched ablations, and completing the QNet runs are the main directions for future work. 

The results presented in this article represent only the performance demonstrated by the model when applied to a total of one layer in the base model. Additionally, the adapted version of QuFex utilized in this project does not take into account multi-layering. In future work, it is expected to obtain results when the quantum layer injection is applied to more than one layer.

\appendix
\section{Appendix}
\subsection{Upstream SimCLR testing at different depths}
\label{sec:appendix}
SimCLR upstream training produces a fingerprint feature representation, using augmented photos during training as ground truth, then calculating representation accuracy with model-predicted photos. Table~\ref{tab3} shows that when quantum computing is added to the network with depth greater than or equal to 2.
\begin{table}
\centering
\caption{Average Training Top-5 Accuracy Comparison between Classical SimCLR and Hybrid SimCLR-QNet, tested with 3 different representation layers.}\label{tab3}
\begin{tabular}{|l|l|l|}\hline

Avg. Top-5 Accuracy ($\%$)&  Classical SimCLR& Hybrid SimCLR-QNet\\
\hline
Repr. Layer with 3 layers&  56.62& 66.96\\\hline
Repr. Layer with 2 layers&  53.65& 65.36\\\hline
Repr. Layer with 1 layer& 62.29& 54.62\\\hline
\end{tabular}
\end{table}
\vspace{-20pt}

\subsection{SimCLR Downstream Verification Accuracy}
Downstream testing of upstream training results has two methods: Method 1 discards the Projection Head and adds a linear network layer after the Encoder, called linear evaluation; Method 2 takes out the upstream Encoder and verifies results through KNN Monitor. Table~\ref{tab4}

\begin{table}
\centering
\caption{Testing Top-5 Accuracy Comparison between Classical SimCLR and Hybrid SimCLR-QNet, using two different evaluation methods}\label{tab4}
\begin{tabular}{|l|l|l|}\hline

Top-5 Accuracy ($\%$)&  Classical SimCLR& Hybrid SimCLR-QNet\\
\hline
Linear Evaluation&  11& 12\\\hline
KNN Monitor&  23& 26\\\hline
\end{tabular}
\end{table}

\vspace{-20pt}
%
%
\bibliographystyle{splncs04}
\bibliography{bibliography}
\end{document}